\documentclass[twocolumn,showpacs,preprintnumbers,amsmath,amssymb]{revtex4}

\usepackage{graphicx}
\usepackage{dcolumn}
\usepackage{bm}

\begin{document}
\renewcommand{\vec}[1]{{\bf #1}}
\newlength{\figwidth}
\setlength{\figwidth}{0.3 \textwidth}
\addtolength{\figwidth}{-0.5 \columnsep}

\preprint{}

\title{Algorithm for Linear Response Functions at Finite Temperatures:\\
Application to ESR spectrum of $s=\frac{1}{2}$ Antiferromagnet Cu benzoate}

\author{Toshiaki Iitaka}
\email{tiitaka@postman.riken.go.jp}
\homepage{http://atlas.riken.go.jp/~iitaka/}
\author{Toshikazu Ebisuzaki}
\affiliation{%
Computational Science Division, \\
RIKEN (The Institute of Physical and Chemical Research) \\
2-1 Hirosawa, Wako, Saitama 351-0198, Japan.
}%

\date{\today}

\begin{abstract}
We introduce an efficient and numerically stable method for calculating linear response functions $\chi(\vec{q},\omega)$ of quantum systems at finite temperatures.  The method is a combination of numerical solution of the time-dependent Schroedinger equation, random vector representation of trace, and Chebyshev polynomial expansion of Boltzmann operator. This method should be very useful for a wide range of strongly correlated quantum systems at finite temperatures. We present an application to the ESR spectrum of $s=\frac{1}{2}$ antiferromagnet Cu benzoate.
\end{abstract}

\pacs{75.10.Jm, 75.40.Mg}
\keywords{linear response,quantum spin,ESR}
\maketitle

\section{\label{sec:level1} Introduction}

Computational physicists often face to problems of calculating the linear response functions or the real-time Green's functions of quantum manybody systems with degree of freedom $N \sim 10^{6} $ or more.
Direct diagonalization of the Hamiltonian matrix is the most naive and powerful method for modest system size $N \le 10^{3} $. However, it becomes prohibitive for larger system because the computational time grows as $N^3$.
Therefore efficient numerical algorithms, such as Quantum Monte Carlo methods, Lanczos Methods, and Kernel Polynomial Methods have been developed and applied to various problems. 

Quantum Monte Carlo methods\cite{Suzuki1977,Linden1992} can generate the Green's functions of very large systems since QMC does not need to store the wave functions. They have been successfully used for evaluating the imaginary-time Green's functions and, therefore, various thermodynamic quantities. However, for evaluating dynamical quantities such as AC conductivity, one has to rely on numerical analytic continuation (e.g. {\em Maximum Entropy Method} \cite{Silver1990} ) to obtain the real-time Green's function from the imaginary-time one. This procedure is not straightforward because the statistical errors are amplified by numerical analytical continuation and the default model for the MEM must be assumed {\it a priori}.

Lanczos Methods\cite{Lanczos1950,Dagotto1994} have been useful techniques for evaluating dynamical responses of relatively large systems. The LM uses Lanczos recursion formula with Matrix Vector Multiplications ($|\phi'\rangle=H|\phi\rangle$) to tridiagonalize the Hamiltonian matrix, which leads to a continued fraction representation of the real-time Green's function. The drawback of LM is numerical instability of Lanczos recursion formula caused by accumulation of roundoff errors for large numbers of MVM's. Recently, LM was extended to finite temperatures (Finite Temperature LM) by introducing random sampling over the ground and excited eigenstates \cite{Jaklic1994}. However, FTLM has a weak point that the number of excited eigenstates to be calculated increases rapidly as temperature or system size becomes large. Therefore reduction of computational costs by exploiting symmetries of the system becomes crucial.

Kernel Polynomial Method \cite{Silver1994,Wang1994} calculates density of states and linear response functions by using Chebyshev polynomial expansion \cite{Kosloff1986,Vijay2002,Recipe} of broadened delta functions.  The Chebyshev polynomials are obtained through Chebyshev recursion formula by using MVM's.
Unlike Lanczos recursion, Chebyshev recursion avoids accumulation of numerical roundoff errors even for large numbers of MVM's. The use of KPM for linear response functions has been limited to the ground state calculation.

Fermi-Weighted Time-Dependent Method  \cite{Iitaka1997} is a combination of conventional time-dependent method \cite{Heller1993,Iitaka1994,Williams1985}, random vector representation of trace \cite{Hams2000}, and Chebyshev polynomial expansion of step function $\theta(E_f-H)$ to extract occupied and unoccupied one-particle states of noninteracting fermi particles in the ground state. This method was successfully applied to optical absorption of silicon nano-crystallites\cite{Nomura1997}.

Our new algorithm, the Boltzmann-Weighted Time-Dependent Method, calculates linear response functions at finite temperatures by using the Boltzmann-weight operator in place of the Fermi-weight operator. The BWTDM has an advantage to FTLM that the computational time does not increase as temperature increases, because BWTDM does not calculate any eigenstates at all. Therefore BWTDM does not need use of symmetries and can easily treat disordered systems.  

In this Letter, we introduce BWTDM and apply it to study Electron Spin Resonance spectrum of one-dimensional $s=\frac{1}{2}$ antiferromagnet Cu benzoate $Cu(C_6H_5COO)_2\cdot3H_2O$, especially the dynamical crossover between {\em spinon} excitation and {\em breather} excitation as a function of temperature.
Nonperturbative analytical calculation of this phenomena is a very difficult problem which still open. Even within Sine Gordon theory, the linear response functions have been studied only at high temperature regime (by perturbation) and at zero temperature. 
We compare our numerical results with preceding experimental and theoretical studies \cite{Oshima1978,Asano2000,Ogasahara2002,Oshikawa1999,Affleck1999,Oshikawa2002,Oshikawa2002b,Oshikawa1997,Essler1998,Essler1999,Lou2002}.

\section{The Model}

The effective Hamiltonian of one-dimensional $s=\frac{1}{2}$ antiferromagnet Cu benzoate may be written \cite{Oshikawa1997,Essler1998,Essler1999} as
\begin{eqnarray}
\label{eq:hamiltonian}
H&=& \sum_{j=1}^{N_{spin}} \left[ J\mathbf{s}_{j}\cdot \mathbf{s}_{j+1}  
-g \mu_B H_x s^x_j +(-1)^j g \mu_B h s^y_j \right].
\end{eqnarray}
The first term is the isotropic Heisenberg antiferromagnet where $\mathbf{s}_{j}$ is spin operator at the $j$-th site, $J/k_B=17.2K=1.48 (meV)$ is the exchange interaction \cite{Oshima1978}, and the sums are taken over $N_{spin}$ spins with periodic boundary conditions. 
The second is the normal Zeeman term due to the external uniform magnetic field $H_x$. 
The third is the staggered Zeeman term due to induced staggered magnetic field $h= 0.065H_x$ originating from the Dzyaloshinskii-Moriya (DM) interaction and the staggered component of the $g$ tensor\cite{Oshikawa2002}. 
In the following we assume that $g=2.25(2.29)$ with $H_x$ along $c$($c''$)-direction. Unit of magnetic field and frequency become  $J/g_{c}\mu_B=11.4(T)$, $J/g_{c''}\mu_B=11.2(T)$, and $J/h=359 (GHz)$, respectively.

According to the linear response theory, absorption of electromagnetic waves of frequency $\omega$ and polarization $\mu$ is expressed as
\begin{equation}
I^\mu(\omega)=\frac{H_R^2\omega}{2} \chi^{''}_{\mu\mu}(q=0,\omega)
\end{equation}
where $H_R$ is the amplitude of the electromagnetic waves and $\chi^{''}_{\mu\mu}(q,\omega)$ is the imaginary part of the dynamical magnetic susceptibility, 
\begin{equation}
\chi^{''}_{\mu\mu}(q,\omega)=
(1-e^{-\beta\omega}) 
\ \mathrm{Im} \lim_{\eta \rightarrow +0}\int_0^\infty dt e^{-i(\omega-i\eta) t} g_{q}^{\mu}(t)
\end{equation}
at inverse temperature $\beta=1/(k_BT)$.
The correlation function is defined by
\begin{equation}
\label{eq:correlation}
g_{q}^{\mu}(t)= \mathrm{Tr} \left[ e^{-\beta H} M^\mu_{-q} e^{+iHt} M^\mu_{+q} e^{-iHt} \right]
/ \mathrm{Tr} \left[ e^{-\beta H} \right]
\end{equation}
where the magnetization operator is
$
M^\mu_{q}=\sum_{j=1}^{N_{spin}}  s^{\mu}_j \frac{e^{iqj}}{\sqrt{N_{spin}}}
.
$

\begin{figure}
\resizebox{\figwidth}{!}{\includegraphics{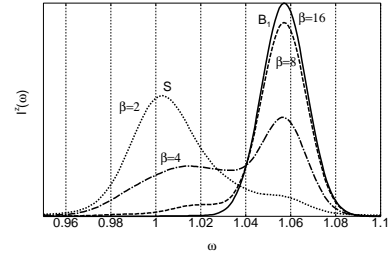}}
\caption{\label{fig1} ESR spectra of normal polarization, $I^z(\omega)$, calculated with  $\beta=2 \sim 16$, $H_x=1.0$, $N_{spin}=16$, $\eta=0.01$ and $N_{rand}=16$. $S$ and $B_1$ stand for spinon excitation and first breather excitation, respectively.}
\end{figure}

\section{Algorithm}
The essence of BWTDM is evaluation of eq.~(\ref{eq:correlation})  by using {\em Boltzmann-weighted random vectors}, $|\Phi_{Boltz}\rangle=e^{-\beta H/2}|\Phi\rangle$, 
\begin{eqnarray}
\label{eq:correlation2}
g_{q}^{\mu}(t)
&=& 
\frac{
\langle\langle \ \langle \Phi_{Boltz}| M^\mu_{-q} e^{+iHt}| M^\mu_{+q} |e^{-iHt} | \Phi_{Boltz} \rangle \ \rangle\rangle
}{ 
\langle\langle \ \langle \Phi_{Boltz}|\Phi_{Boltz} \rangle \ \rangle\rangle
}
.
\end{eqnarray}
In the above equation, the {\it uniform random vector}, $|\Phi \rangle$, is defined by 
\begin{equation}
\label{eq:randomvec}
|\Phi \rangle \equiv \sum_{n=1}^N |n \rangle \xi_n 
\end{equation}
where $\{ |n \rangle \}$ is the basis set composed of eigenvectors of $s^z_j$'s, and  $\xi_n$ are 
a set of complex random variables  satisfying
\begin{equation}
\label{eq:statis}
\left\langle \left\langle \  \xi_{n'}^*  \xi_{n} \  \right\rangle \right\rangle = \delta_{n'n}
.
\end{equation}
Here $\left\langle \left\langle \   \cdot \  \right\rangle \right\rangle \ $ stands for the statistical average. 
Then $\left\langle \left\langle \ \langle \Phi | X | \Phi \rangle \ \right\rangle \right\rangle $ gives the trace of $X$,
\begin{eqnarray}
\label{eq:trace.monte}
\lefteqn{
\left\langle \left\langle \   \langle \Phi | X | \Phi \rangle \  \right\rangle \right\rangle \  
= \sum_{n}
\left\langle \left\langle \ \xi_{n}^* \xi_n  \  \right\rangle \right\rangle
 \langle n|X| n \rangle   
}
\nonumber \\
&&
+\sum_{ n \ne n'} \left\langle \left\langle \ \xi_{n'}^* \xi_n  \  \right\rangle \right\rangle \  \langle n'|X| n \rangle \nonumber \\
&=& \sum_{ n } \langle n |X| n \rangle  
= {\rm tr} \left[ X \right]
\end{eqnarray}
where we used eq.~(\ref{eq:statis}).
The Boltzmann operator is evaluated by the Chebyshev polynomial expansion \cite{Kosloff1986,Vijay2002,Recipe},
\begin{equation}
\label{eq:cheby}
e^{-\tau H} |\phi\rangle =\sum_k a_k(\tau) T_k(H) |\phi\rangle
\end{equation}
where $a_k(\tau)$ is the coefficient of the Chebyshev polynomial expansion, and $|\phi\rangle$ is an arbitrary vector. The Chebyshev polynomial of k-th order  $T_k(x)$ is calculated by the Chebyshev recursion formula,\begin{equation}
T_{k+1}(H)|\phi \rangle= 2 H T_{k}(H)|\phi\rangle -T_{k-1}(H)|\phi\rangle 
.
\end{equation}
The time-evolution operator $|\phi; t+\Delta t\rangle =e^{-iH\Delta t}|\phi;t\rangle$ can be evaluated with numerical stability by the leap frog method \cite{Iitaka1994},
\begin{equation}
\label{eq:frog}
|\phi;t+\Delta t \rangle=-2iH\Delta t |\phi;t\rangle +|\phi;t-\Delta  \rangle 
.
\end{equation}

In real calculation, we first generate $N$ complex random numbers $\xi_n$ and construct $|\Phi\rangle$ according to eq.~(\ref{eq:randomvec}). The $|\Phi_{Boltz}\rangle$ can be computed with numerical stability by applying eq.~(\ref{eq:cheby}) repeatedly to $|\Phi\rangle$. Then $\langle \Phi_{Boltz}|\Phi_{Boltz} \rangle$ gives a sample of the denominator of eq.~(\ref{eq:correlation2}).  Here, we introduce another vector 
$
|\Phi_{M_{+q}^\mu}\rangle=M_{+q}^\mu |\Phi_{Boltz}\rangle
$ 
and calculate the time evolution of $|\Phi_{Boltz}\rangle$ and $|\Phi_{M_{+q}^\mu}\rangle$ by eq.(\ref{eq:frog}). At each time $t=n\Delta t$, the matrix element $\langle \Phi_{M_{+q}^\mu};t|M_{+q}^\mu|\Phi_{Boltz};t\rangle$ gives a sample of numerator of eq.~(\ref{eq:correlation2}). 
After calculating the denominator and numerator for $N_{rand}$ random vectors, the ratio of their averages gives $g_q^\mu(t)$ of eq.~(\ref{eq:correlation2}).
Then $\chi^{''}_{\mu\mu}(q,\omega)$ with a finite frequency resolution $\eta$ is calculated by using Gaussian filter,
\begin{equation}
\chi^{''}_{\mu\mu}(q,\omega)= 
(1-e^{-\beta\omega}) 
\ \mathrm{Im} \int_0^{T_{max}} \!\!\!\!\! dt e^{-i\omega t}  g_{q}^{\mu}(t) \left[ A e^{-\eta^2t^2/2} \right]
\end{equation}
where $T_{max}$ is related to $\eta$ by $T_{max} \sim \eta^{-1}$.  In order to avoid finite size effect of spinon excitations, $T_{max}$ is chosen so that $T_{max} < N_{spin}/v_{spin}$ where $v_{spin} \sim 1$ is the spin wave velocity \cite{Essler1999}. This limits the finest frequency resolution. Note that this is not the limit due to the algorithm but due to the finite size model. Much finer resolution can be used for breather modes, which are spatially localized.

Two more numerical details: First, the fluctuation in the result is inversely proportional to the square root of the number of eigenstates participating in $|\Phi_{Boltz}\rangle$ multiplied by $N_{rand}$\cite{Hams2000}.  Therefore systems at lower temperatures require more random vectors for the same quality of the result. 
Second, since the Chebyshev polynomials are defined only in the range $[-1,+1]$, the Hamiltonian should be rescaled when eq.~(\ref{eq:cheby}) is applied, so that all eigenvalues of $H$ lie in the range $[-1,+1]$.

\begin{figure}
\resizebox{\figwidth}{!}{\includegraphics{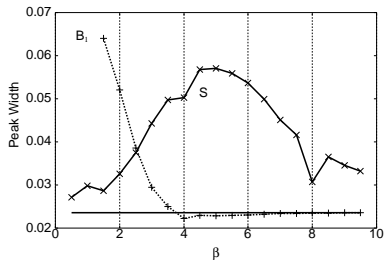}}
\caption{\label{fig2a} (a) Peak width (FWHM) of spinon ($S$) and first breather ($B_1$) excitations calculated with  $\beta=0.5 \sim 9.5$, $H_x=1.0$, $N_{spin}=16$, $\eta=0.01$ and $N_{rand}=16$. The horizontal solid line represents the width due to the finite resolution $\eta=0.01$.}
\addtocounter{figure}{-1}
\end{figure}

\begin{figure}
\resizebox{\figwidth}{!}{\includegraphics{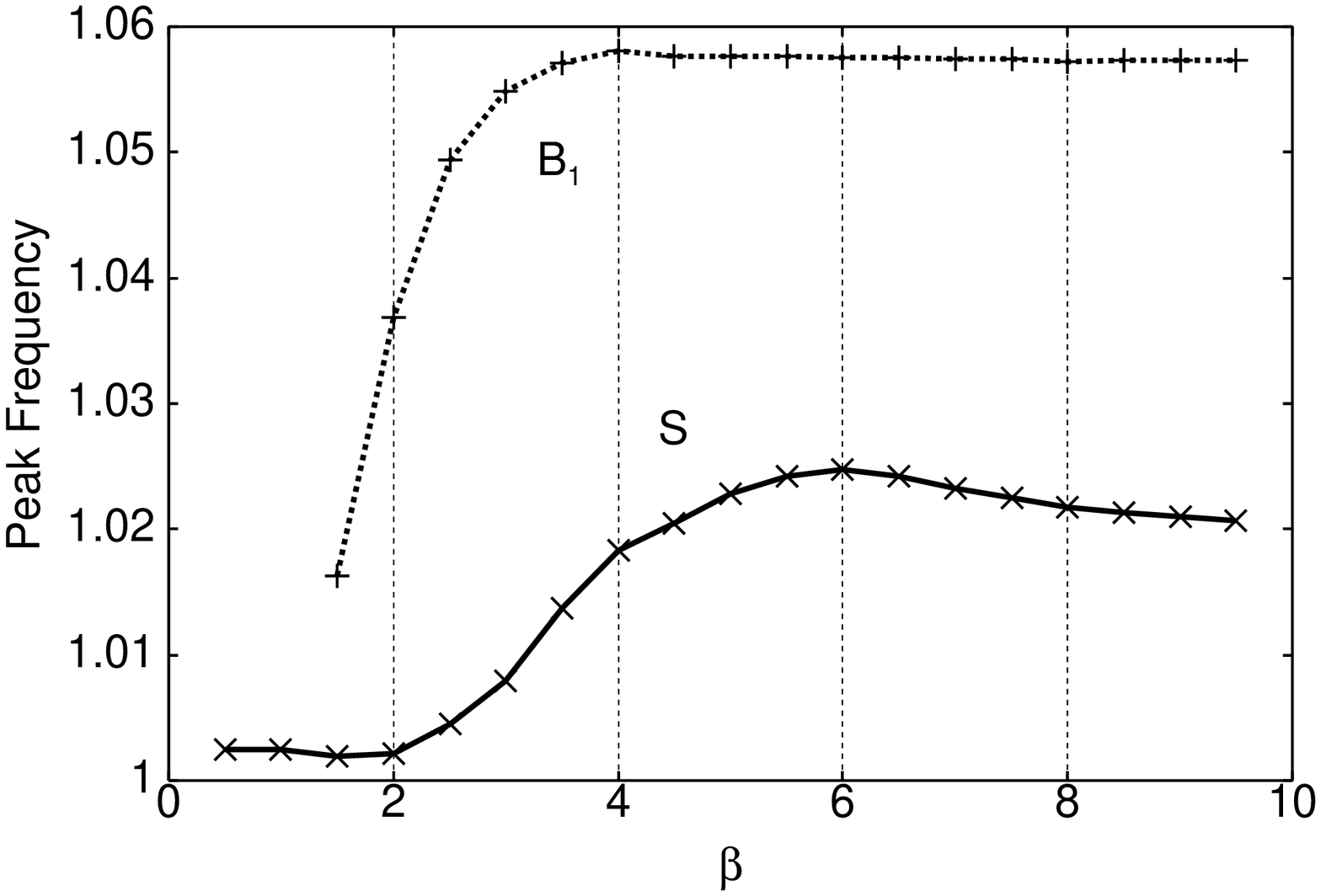}}
\caption{\label{fig2b} (b) Peak frequency of spinon ($S$) and first breather ($B_1$) excitations calculated with  $\beta=0.5 \sim 9.5$, $H_x=1.0$, $N_{spin}=16$, $\eta=0.01$ and $N_{rand}=16$.}
\end{figure}

\section{Results and Discussions}

Figure 1 shows the ESR spectra of normal polarization (Faraday configuration) $I^z(\omega)$ calculated with various temperatures, and Fig. 2 shows the width and peak frequency of the spinon excitation  $S$ and the first breather excitation $B_1$ obtained by fitting two Gaussian peaks to the calculated spectra with the least square method.
At high temperatures ($\beta \le 5$), $S$ is outstanding in the spectrum, and its peak shifts to higher frequency and becomes broadened as temperature decreases while $B_1$ becomes narrower. At low temperatures ($\beta \ge 5$),  $B_1$ prevails while its peak frequency is almost constant and its width becomes much smaller than the numerical resolution $\eta$.   
 This crossover behavior between spinon excitation and first breather excitation has become computable by the invention of BWTDM, and the result is consistent with both experimental \cite{Asano2000} and field-theoretical results \cite{Oshikawa2002}. 
In addition, we can observe in Fig.2 the shift of $B_1$ at high temperatures and the narrowing of $S$ at low temperatures.
We believe these tendencies are real physics but not numerical artifact though further experimental and theoretical studies are necessary to confirm it.
Several other weak peaks appear in our results as well as in experimental results, which are supposed to be higher-order breathers and transitions between excited states. However, we are not going to analyze them here. 

Figure 3 shows the induced energy gap $E_g(H_x)$ calculated from the $B_1$ frequency $\omega$ by solving \cite{Oshikawa1999}
\begin{equation}
\label{eq:nu}
\hbar \omega = \sqrt{(g\mu_BH_x)^2+\left(E_g(H_x)\right)^2}
.
\end{equation}
For $H < 1$, $E_g(H_x)$ scales to $H_x^{2/3}$ as predicted by Oshikawa and Affleck\cite{Oshikawa1997}. 
For $1< H_x <4$, our result seems to deviate from the $H_x^{2/3}$-scaling and have a dip around $H_x^{2/3} \sim 2$. This seems consistent with the experimental results and the density-matrix renormalization-group study \cite{Lou2002}. At such a high field, assumption of Sine Gordon theory, that the $H_x$ is weak and the frequency is small, is probably broken.
Since the gap is calculated as a small difference of two large numbers, $\hbar \omega$ and $g\mu_BH_x$, the agreement of our result with the experimental value and the $H_x^{2/3}$-scaling at low fields demonstrates the numerical accuracy of our algorithm.

\begin{figure}
\resizebox{0.8\figwidth}{!}{\includegraphics{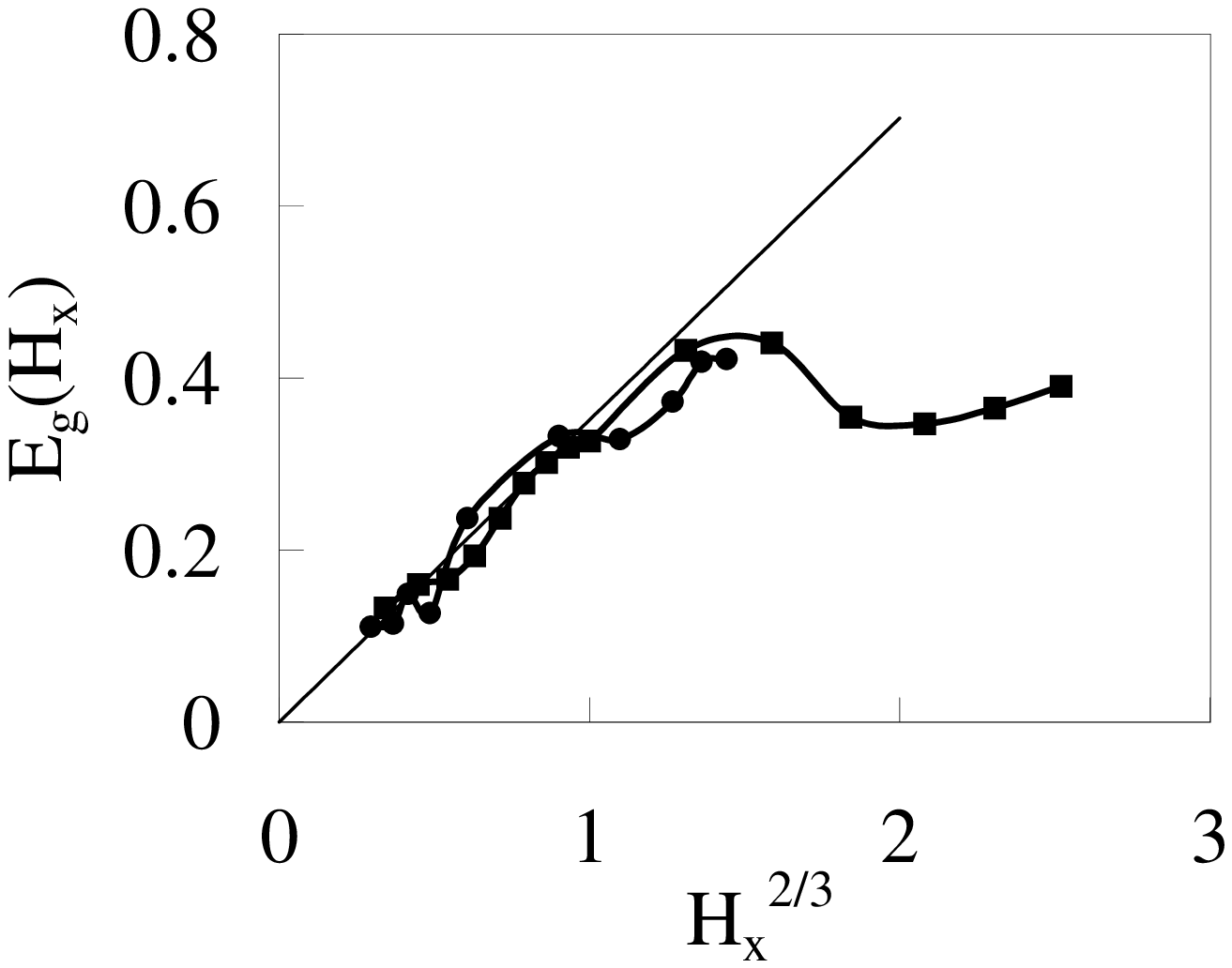}}
\caption{\label{fig3} 
Solid curve with filled squares shows $E_g(H_x)$ calculated from our results with $\beta=16$, $N_{spin}=16$, $\eta=0.01$ and $N_{rand}=16$.
Solid curve with filled circles shows $E_g(H_x)$ for $H_x \parallel \mathrm{c-axis}$ calculated from the experimental $B_1$ peak read from Figure 4 of Ref.~\cite{Asano2000}.
Thin line shows a fit $\omega=c H_x^{2/3}$ to the experimental result. \cite{Asano2000}.
}
\end{figure}

In summary, we developed an efficient and stable algorithm for linear response functions at finite temperatures, and studied the ESR spectra of $s=\frac{1}{2}$ antiferromagnet Cu benzoate for a wide range of temperature and magnetic field. We reproduced experimental results of the spinon-breather crossover as a function of temperature. Temperature dependence of the width and shift of the peaks are also calculated, which are consistent with experiments and analytical theories. The calculated frequency of $B_1$ as a function of $H_x$ agrees well with both experimental and field-theoretical results at low fields, and reproduces the deviation of experimental results from the Sine Gordon theory at high fields, where the low energy assumption of the Sine Gordon theory may be broken and the choice of the compactification radius $R$ becomes ambiguous\cite{Oshikawa2002b}. The advantage of BWTDM is being applicable to finite temperatures, strong magnetic field and high frequency while its weak point is finite size effects ($N_{spin} \leq 20$). The computational cost of BWTDM is moderate. Calculation of one curve in Fig.~1, for example, requires approximately 30 minutes with 8 CPU's of Fujitsu VPP5000 vector-parallel computer. We hope that this algorithm stimulates further numerical investigation of dynamical properties of quantum manybody systems at finite temperatures in general. 

\begin{acknowledgments}
One of the authors (TI) would like to thank Professor Seiji Miyashita for introducing the problem of ESR spectrum of Cu benzoate and Professor Masuo Suzuki for his continuous encouragement.
The results presented here were computed by using supercomputers at RIKEN and NIG.
\end{acknowledgments}

\end{document}